\documentclass[12pt]{article}
\usepackage{graphicx}

\newcommand\pubnumber{KCL-PH-TH/2016-67}
\newcommand\pubdate{\today}

\def\institute{Theoretical Particle Physics \& Cosmology Group, Department of Physics,\\
KingÕs College London, Strand, London WC2R 2LS, United Kingdom}
\def\support{\footnote{Work supported by the UK STFC via the research grant ST/L000326/1.}}

\def\Title#1{\begin{center} {\Large #1 } \end{center}}
\def\Author#1{\begin{center}{ \sc #1} \end{center}}
\def\Address#1{\begin{center}{ \it #1} \end{center}}

\newcommand\pubblock{\rightline{\begin{tabular}{l} \pubnumber\\
         \pubdate  \end{tabular}}}
\newenvironment{Abstract}{\begin{quotation}  }{\end{quotation}}
\newenvironment{Presented}{\begin{quotation} \begin{center} 
             PRESENTED AT\end{center}\bigskip 
      \begin{center}\begin{large}}{\end{large}\end{center} \end{quotation}}
\def\Acknowledgements{\bigskip  \bigskip \begin{center} \begin{large}
             \bf ACKNOWLEDGEMENTS \end{large}\end{center}}

\def\beq{\begin{equation}}
\def\eeq#1{\label{#1}\end{equation}}
\def\eeqn{\end{equation}}

\def\beqa{\begin{eqnarray}}
\def\eeqa#1{\label{#1}\end{eqnarray}}
\def\eeqan{\end{eqnarray}}

\let\bar=\overbar

\def\Dslash{\not{\hbox{\kern-4pt $D$}}}
\def\dslash{\not{\hbox{\kern-2pt $\del$}}}

\def\msb{{\bar{\ssstyle M \kern -1pt S}}}

\begin{document}
\begin{titlepage}
\pubblock

\vfill
\Title{Searching for New Physics in Scalar Top-Pair Resonance}
\vfill
\Author{ J\'er\'emie Quevillon\support}
\Address{\institute}
\vfill
\begin{Abstract}
We consider the effects in the production via gluon fusion in LHC collisions of one or two spin-zero new resonant particles $\Phi$ that decay into a top quark pair.
We revisit previous analyses of the interferences between the heavy-fermion loop-induced $gg \rightarrow \Phi \rightarrow t\bar{t}$ signal and the continuum QCD background $gg \rightarrow t\bar{t}$. 
We show that in the presence of standard model fermions only in the $gg \rightarrow \Phi$ loops, the interference effect is destructive causing a dip in the $t \bar t$ mass distribution. Including New Physics such as additional vector--like quarks leads to a totally different picture as the line-shape can be distorted by peaks and dips. For the time being, the absence of such effects in ATLAS and CMS data constrain models of the production and decays of the $\Phi$ state(s).
\end{Abstract}
\vfill
\begin{Presented}
$9^{th}$ International Workshop on Top Quark Physics\\
Olomouc, Czech Republic,  September 19--23, 2016
\end{Presented}
\vfill
\end{titlepage}
\def\thefootnote{\fnsymbol{footnote}}
\setcounter{footnote}{0}

\section{Introduction}

The $t \bar{t}$ final state has significant continuum background, which presents opportunities as well as problems. Interference effects on the line-shape of the top quark-antiquark invariant mass distribution may be able to provide information on both the real and imaginary parts of the loops involved in the $gg \rightarrow \Phi \rightarrow t \bar{t}$ amplitudes, providing supplementary constraints on the properties of one or two new state(s)~\cite{Djouadi:2016ack}. 

There have been pioneering analyses of possible interference effects in the decays of a spin-zero resonance into a top quark pair, in both the Standard Model~\cite{Htt0} and a   two--Higgs doublet model \cite{Htt1}. 

As it is known, if the $gg \rightarrow \Phi$ cross section is generated by the top quark loops only, the interference is destructive with the net effect of a dip in the measured $t\bar{t}$ cross section beyond the nominal position of the resonance peak. In contrast, we find that if additional heavy quarks contribute to the production amplitude, the interference can become more involved being destructive before and constructive after the mass peaks, for example. The magnitudes of these dips and peaks, which could be measured experimentally, depend on the masses and couplings of the new spin-zero resonances and also potentially to the new particles involved in the initial loop $gg \rightarrow \Phi$.

The ATLAS and CMS collaborations have released analyses of $t\bar{t}$ production at the LHC at 8~TeV, 13 TeV \cite{ATLAS-tt,CMS-tt} give no indication of any structure up to $\sim 3.5$~TeV, setting limits on any downward or upward deviations of the differential cross sections from the background that can be used to constrain the properties of possible new mediating particles. Since $\Phi \rightarrow  t \bar{t}$ decay is the dominant mode in many scenarios Beyond the Standard Model as the low $\tan\beta$ region of the MSSM~\cite{lowtanbeta}, future LHC data should allow any new state to be observed in this channel, and these genuine interference effects should be incorporated in order to properly interpret any New Physics signal, or its absence.

\section{Interference in $\mathbf{gg \rightarrow \Phi \rightarrow t\bar{t}}$}

\begin{figure}[h]
\centering
\includegraphics[height=1.0in]{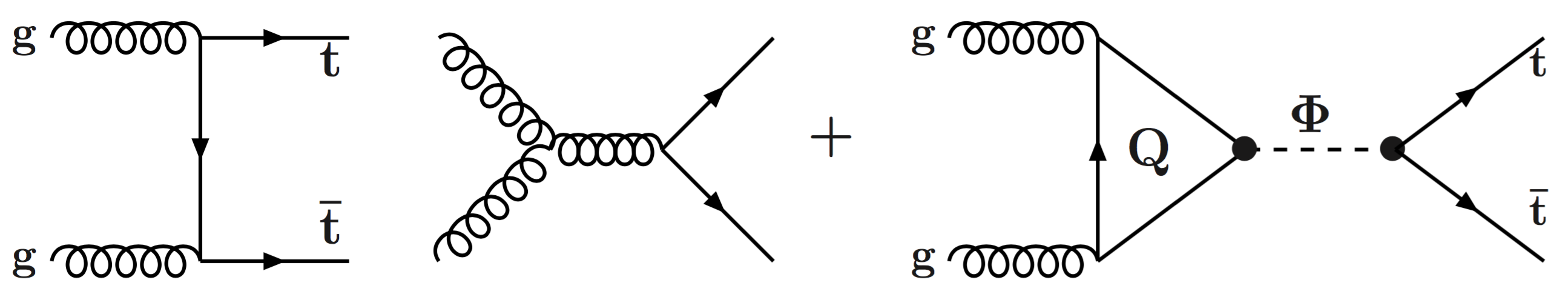}
\caption{Leading-order Feynman diagrams for the continuum QCD background (left) and the resonant $\Phi$ signal (right) in the process $gg \rightarrow \Phi \rightarrow t\bar{t}$.} 
\label{fig:feynmantt}
\end{figure}
The leading-order Feynman diagrams for the QCD background $gg \rightarrow t\bar{t}$ and the signal $gg \rightarrow \Phi \rightarrow t\bar{t}$ are shown in Fig.~\ref{fig:feynmantt}. 
The helicity amplitude of the $gg (\rightarrow \Phi ) \rightarrow t\bar{t}$ process, when the contributions of the continuum backgrounds and a single resonant signal process are added, is given by 
\vspace{-0.4cm}
\beqa
{\cal A}^{\Phi }_{gg \rightarrow t\bar{t}} =  {\cal A}^{\Phi }_{bck} + {\cal A}^{\Phi }_{sig}, \qquad   with \quad {\cal A}^{\Phi}_{sig} = {\cal A}_{sig} e^{i\phi_{sig}} \frac{M_\Phi^2}{\hat{s} -M_\Phi^2 +i M_\Phi \Gamma_\Phi} \, , \nonumber
\eeqan
where the expression of the helicity amplitude of the QCD background (real number) ${\cal A}^{\Phi }_{bck}$ can be find in~\cite{Htt0}. Concerning the helicity amplitude of the signal, ${\cal A}^{\Phi}_{sig}$, we factor out the Breit-Wigner propagator of the resonance amplitude. 
${\cal A}_{sig}$ and $\phi_{sig}$ are respectively the magnitude and the phase of the signal amplitude. The signal amplitude can be decomposed into the gluon fusion production of the $\Phi$ state and its tree-level decay, ${\cal A}_{sig} e^{i\phi_{sig}} = {\cal A}_{gg\Phi} {\cal A}_{\Phi tt} $. We then realize that the complex phase for the signal, $\phi_{sig}$, can be generated by loop diagrams involved in the process $gg \rightarrow \Phi$. In order to compute the cross section we will have to square the total helicity amplitude, $|{\cal A}^{\Phi }_{gg \rightarrow t\bar{t}}|^2 =  {{\cal A}^{\Phi }_{bck}}^2 + |{\cal A}^{\Phi }_{sig}|^2 + {\cal A}^{\Phi}_{bck} \times 2 \Re({\cal A}^{\Phi}_{sig})$. The second term is the usual Breit-Wigner (peak) contribution and the last term is the so called interference contribution which is most of the time negligible and then disregarded in many phenomenological analyses. However in our case study, this contribution leads to dramatic experimental signatures which could allow us to discover New Physics.

\begin{figure}[!h]
\centerline{ 
\includegraphics[scale=0.25]{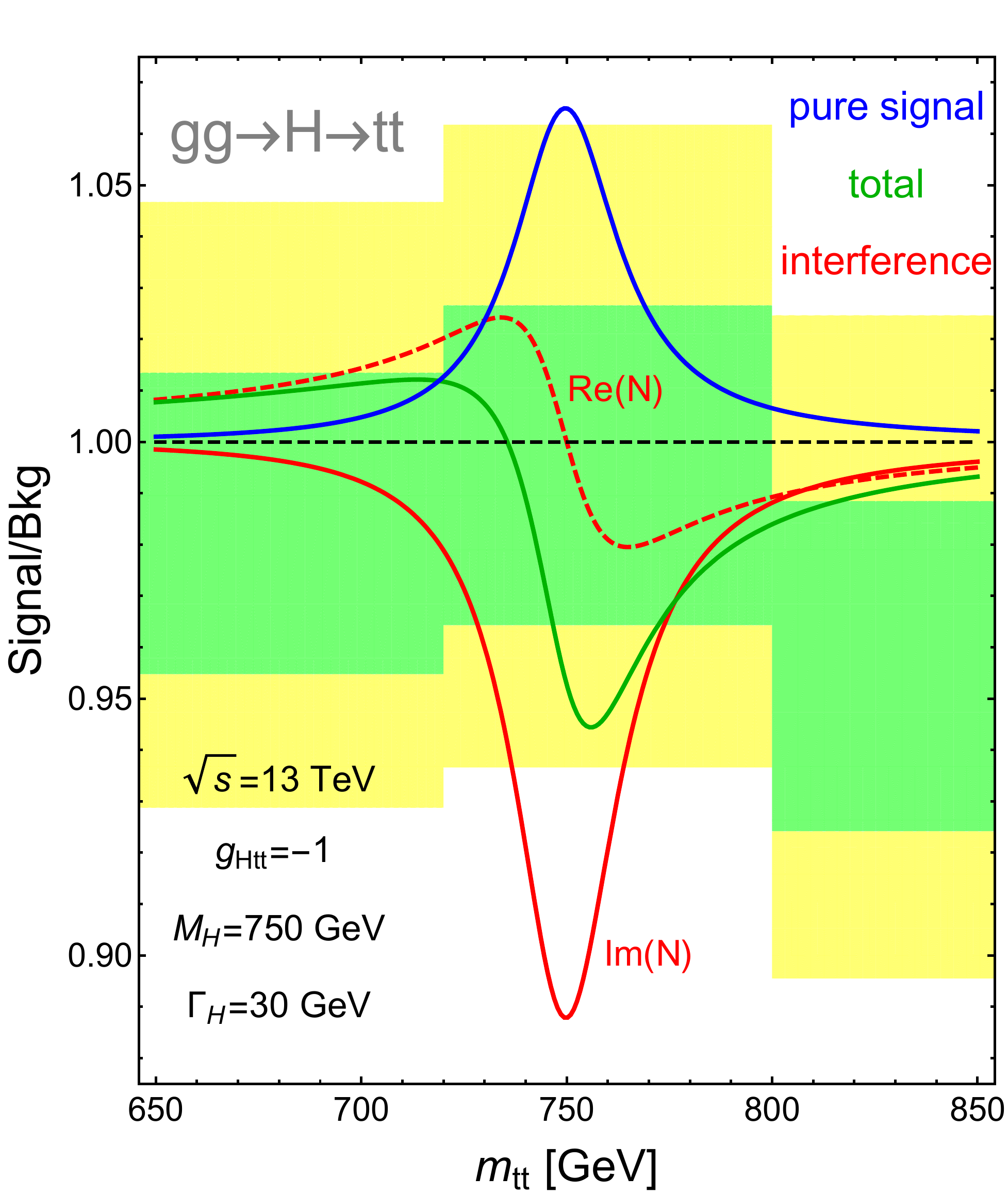} \hspace*{1.5cm}
\includegraphics[scale=0.25]{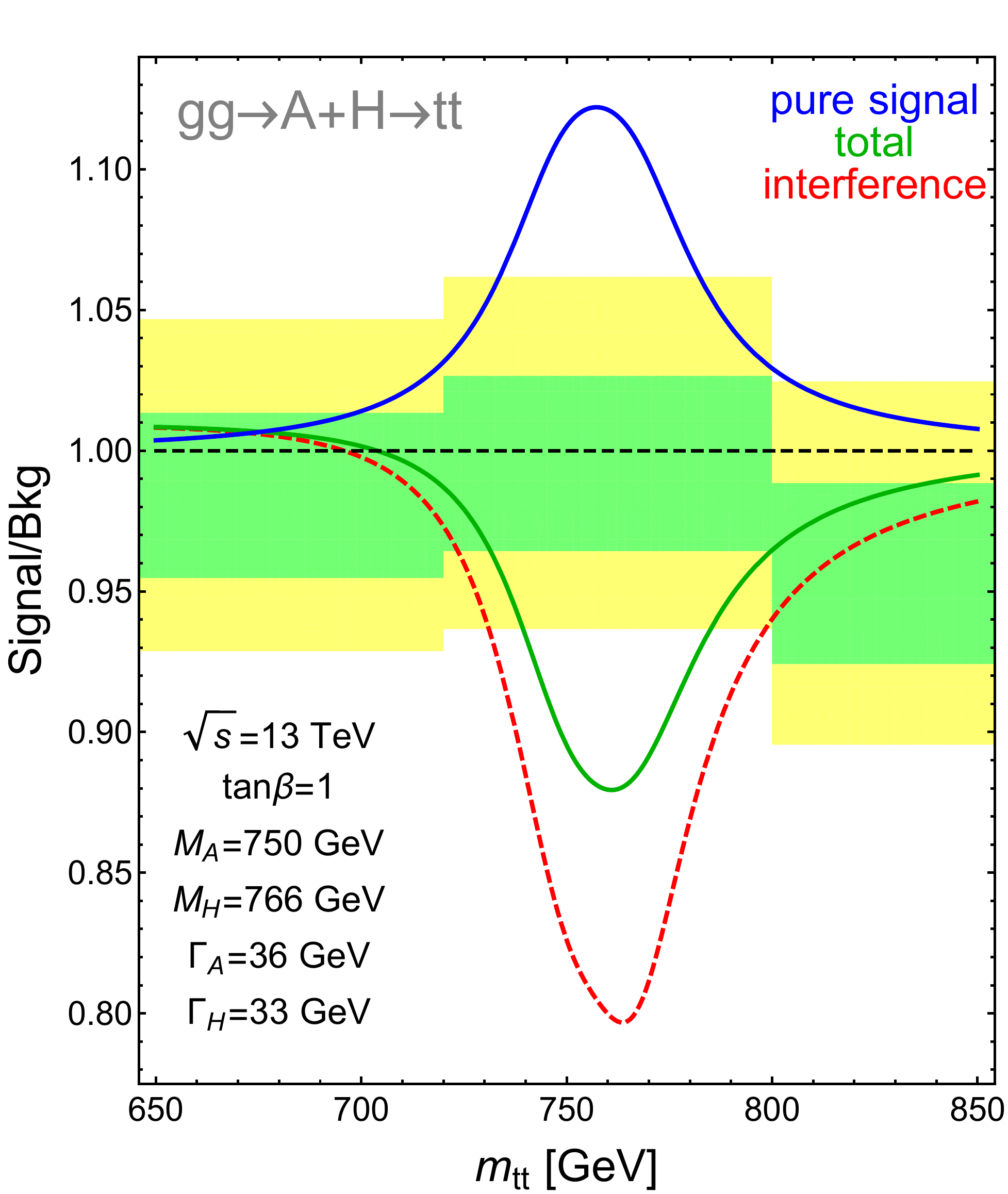}} 
\caption{The contributions to the line-shapes of a CP--even $H \to t {\bar t}$ with mass 750~GeV and total width $\Gamma_H = 30$~GeV (left panel) and the contributions to the combined $t \bar t$ line-shape of a CP--odd $A$ with mass 750~GeV and 36~GeV width and a CP--even $H$ with mass 766~GeV and 33~GeV width, with 2HDM  couplings,  neglecting interference (solid blue line), the contribution of interference (dashed red line) and the combination of the two (solid green line).}
\label{fig1}
\end{figure}

In the case where the loop factor ${\cal A}_{gg\Phi}$ does not acquire any phase (which corresponds to heavy fields circulating inside the loop regarding the resonance mass) then the interference term take the form $2 {\cal A}^{\Phi}_{bck} {\cal A}_{sig} \frac{M^2(\hat{s}-M^2)}{(\hat{s}-M^2)^2+M^2\Gamma^2}$, where $M, \Gamma$ are the mass and width of the new resonance and $\hat{s}$ is the partonic center of mass energy. We notice that when the resonance is produced on-shell, there is no interference effect and that this contribution is entirely antisymmetric around M (regarding $\hat{s}$), which means that the total cross section is not altered by such effect. Such contribution can be visualized by the dashed red line in Fig.~\ref{fig1}.

In the case where the loop factor ${\cal A}_{gg\Phi}$ is a complex number (which corresponds to light fields circulating within the loop such as the top quark of the Standard Model) then the interference term take the form \\
$2 {\cal A}^{\Phi}_{bck} {\cal A}_{sig} \left( \cos\phi_{sig} \frac{M^2(\hat{s}-M^2)}{(\hat{s}-M^2)^2+M^2\Gamma^2}  + \sin\phi_{sig} \frac{M\Gamma}{(\hat{s}-M^2)^2+M^2\Gamma^2} \right)$, where the first term comes from the real part of the propagator as the previous case, however the second term is a new contribution which comes from the imaginary part of the resonance propagator. We notice that this new contribution does not vanish on-shell and does contribute to the total cross section. Such effect can be visualized by the solid red line in Fig.~\ref{fig1}.

Consequently, we realize how the line-shape of the top quark pair invariant mass distribution might be distorted by New Physics (namely the mass, width and couplings of the potentially new resonance but also the mass and couplings of the potentially new fields contributing to the loop induced coupling). 

The left panel of Fig.~\ref{fig1} shows the results for a singlet scalar $H$ for which $\Gamma_H \approx \Gamma( H \to t\bar t) = 30$~GeV with a unit-normalized coupling $g_{H t {\bar t}} = -1$ (the minus sign is defined relative to the sign of the standard $h t {\bar{t}}$ coupling). We display separately the interference in the real part (dashed red line) and the interference in the imaginary part of the production amplitude (solid red line), as well as the line-shape without any interferences (solid blue line) and with interferences (solid green line).

The interference coming from the real part of the propagator changes sign across the nominal $H$ mass, whereas the interference coming from the imaginary part (which is only due to the top quark loop in $gg \to H$ production here) is larger in magnitude and always negative. This is why, the combined interference effect overwhelms the putative peak and is negative, resulting finally in a $m_{t\bar{t}}$ distribution with a dip. The depths of the dip almost reach the ATLAS 2-$\sigma$ lower limit. However, when integrated over the ATLAS bin the net effect would be $< 1 \sigma$. We note that the dip is not symmetric about the resonance mass, M, and greater sensitivity to interference effects could be obtained by comparing off-centre bins $[M - X, M]$~GeV and $[M, M + X]$~GeV, where the choice of $X$ depends on the attainable mass resolution. However, the dip structure in the very small width case (~1 GeV) is unlikely to be unobservable because of the poor resolution in $m_{t\bar{t}}$.

In all the Figures we show the results of calculations of the ratios \\
\centerline{ (S + B)/B = (signal + background)/(background alone),} \\
for several scenarios. Both CMS and ATLAS have published measurements of the $t \bar{t}$ cross section as a function of $m_{t\bar{t}}$, providing also values of the ratio of the data to smoothed fits to the background (the ATLAS 8-TeV data~\cite{ATLAS-tt} are more constraining, so we focus on them). The result is displayed in our plots as 1- and 2-$\sigma$ green and yellow bands. The data were used in~\cite{ATLAS-tt} to present upper limits on peaks above the background. However, we show that the data could be more significant for the constraints they impose on dips below the background level.

We display on the right panel of Fig.~\ref{fig1} the combined effects in the 2HDM with a resonance masses of 750~GeV for the pseudoscalar $A$ and 766~GeV for the scalar $H$ and corresponding decay widths $\Gamma_A \approx \Gamma (A \rightarrow t\bar{t}) = 36$~GeV and $\Gamma_H \approx \Gamma (H \rightarrow t\bar{t}) = 33$~GeV. The solid blue line is the result that would be obtained neglecting any interference effects, the dashed red line is the contribution of the interference term, and the solid green curve is the combination of both. We assume here that  the top quark only contributes to the $gg \rightarrow H, A$ production amplitudes. We see that the interference effects in this case cause a dip that is presumably excluded by the ATLAS 8-TeV data at the 2-$\sigma$ level.

\section{New Physics in the Loop}

As already mentioned, the gluon-gluon-$\Phi$ vertex involves the famous loop functions (the analytic expressions and various plots can be find in \cite{Djouadi:2016ack}).
As it is well known, they posses a characteristic jump at the energy threshold $\sqrt{\hat{s}} \simeq 2 m_{f}$, where $m_{f}$ is the mass of the particle running in the loop (the top for example as we considered so far and eventually extra vector-like quark or supersymmetric top quark as we will study in the following). As an example, for heavy fermions, whose mass satisfy $m_{f} \gg \sqrt{\hat{s}}/2$, the loop function is entirely real, almost constant and non vanishing. This leads to the well known fact that any heavy chiral fermion which acquires its mass through the Higgs mechanism will contribute to the gluon-Higgs coupling by a constant amount $\propto \alpha_{s}/v$ (which is the reason why a fourth fermion generation is currently excluded).
When the partonic center of mass energy crosses the energy threshold, $2 m_{f}$, the loop fonction develops an imaginary part which will add a new contribution to the interference terms as we detailed before.
This new phase will not only induce the imaginary part of the propagator to contribute to the total cross section but it will also modulate the real and imaginary contributions, practically speaking, it will amplify or diminish the peak and dip structure against the dip only one.
In the scenario, $\phi_{sig} = \pi/2$, only the interference effect due to the imaginary part of the propagator exists and then the line-shape of the $m_{t\bar{t}}$ distribution has a pure dip interference structure. This corresponds to a scalar resonance mass of $\sim 1.2$~TeV or a pseudo scalar mass of $\sim 850$~GeV.
On the contrary, in the scenario, $\phi_{sig} = \pi/4$, the interference effects due to the real part and the imaginary part of the propagator are of comparable strength and then the line-shape of the $m_{t\bar{t}}$ distribution have a subtil mix between a peak and dip interference structure and a pure dip effect. This corresponds to a scalar resonance mass of $\sim 550$~GeV or a pseudo scalar mass of $\sim 450$~GeV.

\begin{figure}[!]
\centerline{ 
\includegraphics[scale=0.25]{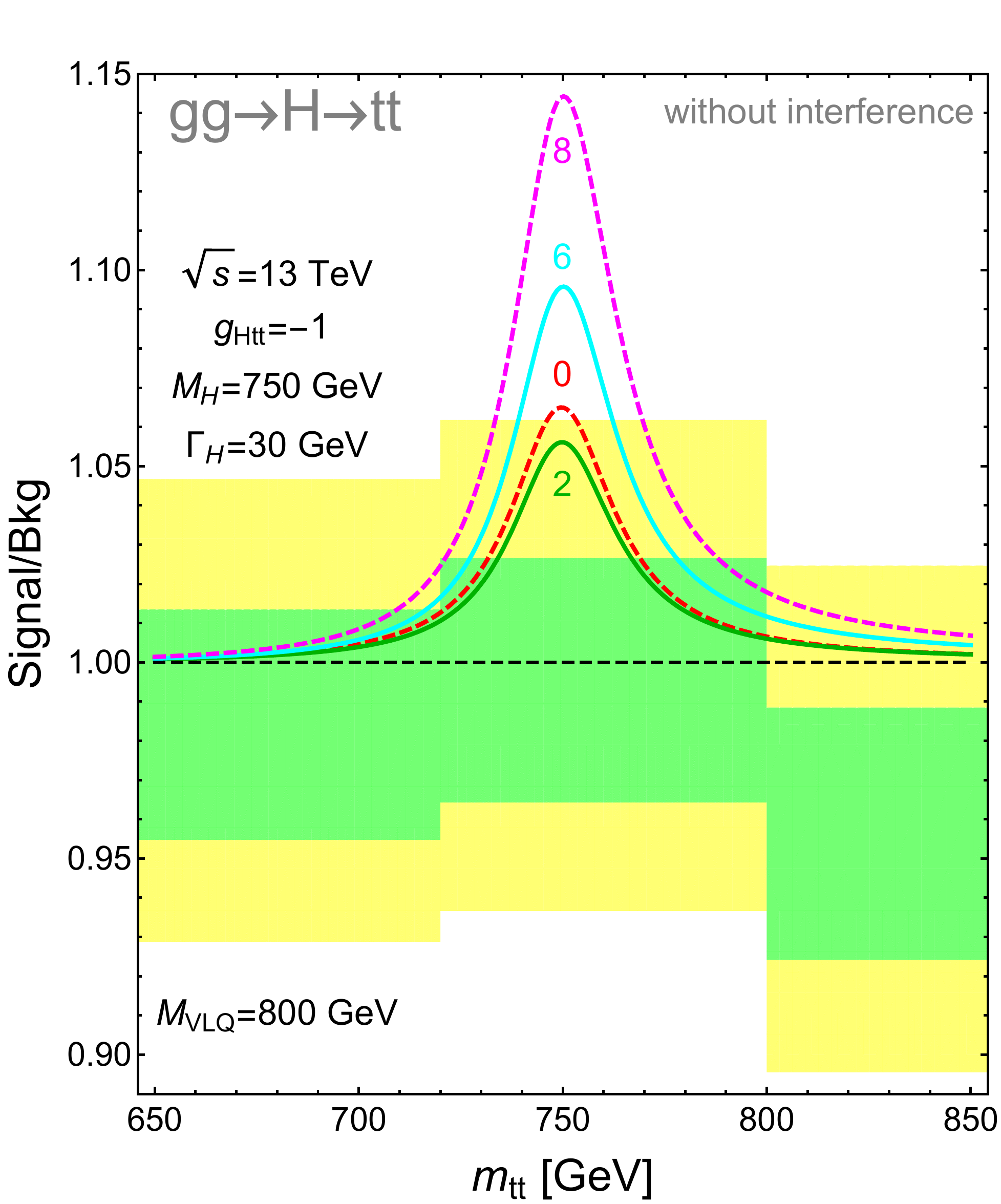} \hspace*{1.5cm}
\includegraphics[scale=0.25]{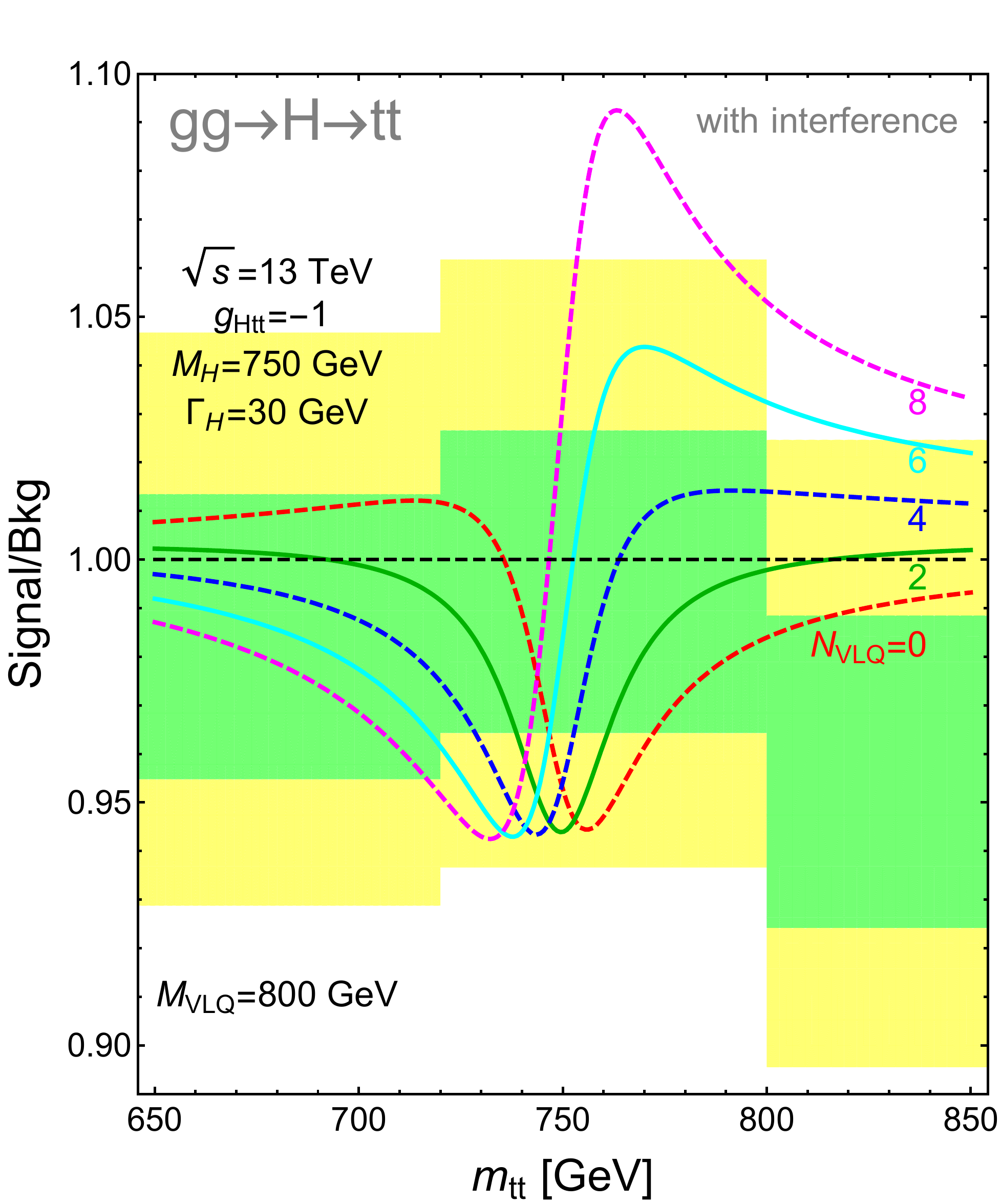}} 
\caption{The contributions to the line-shapes of a CP--even $H \to t {\bar t}$ with mass $750$~GeV and total width $30$~GeV, showing the effects of varying numbers of vector--like quarks with masses~$800$~GeV. The left panels neglect interferences, which are included in the right panels.} 
\label{fig2}
\end{figure}

Vector-like quarks are a perfect example to visualize the modification of this phase due to an imaginary loop factor. Contrary to chiral fermions, vector-like quarks would decouple and then do not contribute anymore to the gluon-$\Phi$ vertex, but as we will see substantial deviation can be obtained for masses around the TeV scale.
Fig.~\ref{fig2} shows the effects of including different numbers of vector-like quarks $Q$ in the loop responsible for $gg \rightarrow H$, assuming $\Gamma_H = 30$~GeV and common $Q$ masses of 800~GeV and universal positive, unit-normalized $HQ{\bar{Q}}$ couplings. In the absence of interference (left panel) we see that adding 6 or 8 such heavy vector-like quarks takes the peak outside the 2-$\sigma$ ATLAS range. However, the right panel reveals a different picture when interference effects are included. There are dips for $N = 0, 2, 4, 6, 8$ vector-like quarks, but there are also significant peaks for $N = 6, 8$, in particular. The net result of integrating over the ATLAS $[720, 800]$~GeV bin would lie within the 2-$\sigma$ range. However we see again the potential gain in sensitivity to the interference coming from the real part of the propagator (antisymmetric interference effect) that could be obtained by using off-centre $[750 - X, 750]$~GeV and $[750, 750 + X]$~GeV bins. 

The interesting scenario where stops would participate to the induced gluon-$\Phi$ coupling has been studied by Carena and Liu \cite{Carena:2016npr}. It has been concluded that in the case of small Left-Right mixing the stop contribution is typically negligible compared to the Standard Model top quark contribution. However in a scenario with a large Left-Right mixing the stop interference effect would dominate the top quark contribution. This would transform the pure dip structure associated to the top quark into a typical peak and dip structure.

\section{Conclusion}
The interference effects are quite complex and their measurement would provide information on both the real and imaginary parts of the $gg \rightarrow \Phi$ amplitude.
Negative interference effect may cause the total cross sections to exhibit a dip instead of a bump, invalidating limits on resonances based on putative bump (Breit-Wigner) signatures. This occurs, for instance, in the case where the production of the $\Phi$ states is initiated by the standard model top quark loops only. The presence of additional vector--like quarks or supersymmetric top quarks might change the situation since peaks followed by dips might occur, whose analysis would require judicious off-centre binning.

\vspace{-0.4cm}
\Acknowledgements
I am grateful to my collaborators Abdelhak Djouadi and John Ellis.

\end{document}